\documentstyle[12pt]{article}

\newcommand{\be}{\beta}
\newcommand{\ga}{\gamma}
\newcommand{\Ga}{\Gamma}
\newcommand{\de}{\delta}
\newcommand{\ep}{\epsilon}
\newcommand{\m}{\mu}
\newcommand{\n}{\nu}
\newcommand{\r}{\rho}
\newcommand{\f}{\varphi}
\newcommand{\Lc}{\cal L}
\newcommand{\Vc}{\cal V}
\newcommand{\da}{\dagger}

\begin{document}
\begin{titlepage}
\hspace{10.3cm}{\small{CALT-68-1913}}

\hspace{10.3cm}{\small{DOE RESEARCH AND}}

\hspace{10.3cm}{\small{DEVELOPMENT REPORT}}
\vskip50pt
\begin{center}
{\Large\bf Nonrenormalization Theorem\\
 \vskip5pt for Gauge Coupling in $2+1D$}\\
\vskip50pt \bf A.N.Kapustin\footnote[2]{Work supported in part by the
 U.S.Dept. of Energy under Grant No. DE-FG03-92-ER40701}\\
\vskip14pt {\it California Institute of Technology\\
Pasadena, CA 91125, U.S.A.}\\
\vskip20pt{\bf P.I.Pronin}\\
\vskip14pt \it Physics Department\\
\it Moscow State University\\
\it 117234 Moscow, Russia
\end{center}
\vskip40pt
\abstract{
We prove that $\be$-function of the gauge coupling in $2+1D$ gauge theory
coupled to any renormalizable system of spinor and scalar fields is zero. This
result holds both when the gauge field action is the Chern-Simons action and
when it is the topologically massive action.}
\end{titlepage}
\baselineskip=17pt
\section{Introduction.}
It is characteristic of $2+1D$ that one can construct a local first-order
action
which is invariant with respect to infinitesimal gauge transformations. This is
the Chern-Simons action
\[S=\frac{k}{4\pi} \int Tr(A dA+\frac{2}{3}A^3).\]
One can easily see that the theory of scalar or spinor fields interacting with
the Chern-Simons field is renormalizable. On spacetime with topology $\cal
R\times M$ where $\cal M$ is simply connected, the Chern-Simons field has no
 physical degrees of freedom and can be expressed through other fields. If we
 add the Maxwell term to the action then the gauge field will have one physical
 degree of freedom describing a massive vector particle. Such gauge field is
 called topologically massive. The theory of the topologically massive gauge
 field interacting  with spinors, or scalars, is also renormalizable. (The
 theory with Maxwell term only is not good since it has terrible infrared
 divergencies~\cite{rao} .)

On general grounds one could expect that $g^2=4\pi/k$ requires infinite
 renormalization, and therefore $g$ experiences a nontrivial renormalization
group (RG) flow driven by the corresponding $\be$-function
\[\frac{dg(t)}{dt} =\be (g).\]
However, there are strong reasons to suspect that in fact $\be (g)$ is zero in
 any order of perturbation theory. The first (trivial) observation is that in
 the nonabelian case the Chern-Simons action is not invariant with respect to
 large gauge transformations. Rather it is shifted by a number proportional to
 the winding number of the gauge transformation~\cite{witten}. To ensure that
 $e^{iS}$ be invariant, $k$ has to be quantized ($k\in\cal Z$  for $G=SU(N)$),
 otherwise the theory will have a global anomaly. Nontrivial renormalization of
 $k$ would spoil the quantization of $k$ imposed on the classical level and
 introduce the global anomaly. However, this argument does not prove that $\be
 =0$, rather it shows that in case $\be \neq 0$, the theory probably will not
 be nonperturbatively consistent. Another argument comes from direct
 perturbative calculations which were performed by several authors. Chen et
 al.,~\cite{chen} showed that $g$ is not renormalized up to two-loop order in
 the  theory of the nonabelian Chern-Simons field interacting with scalars or
 spinors. Avdeev et al.,~\cite{avdeev} calculated two-loop RG-functions in the
 most general abelian theory including scalars and spinors interacting with
each
 other and with the Chern-Simons field. They also found that $\be (g)=0$.

As for exact results, Delduc et al.,~\cite{delduc}
 proved that $g$ is not renormalized in any order of loop expansion in the
 pure Chern-Simons theory (without matter fields). Moreover, they showed that
 the anomalous dimension of the gauge field also vanishes, so the theory is
finite; the latter fact is not true if matter fields are present~\cite{chen}.
One may say that the pure Chern-Simons theory is not a very good example of the
generic behaviour since it has no physical degrees of freedom and in the axial
gauge is reduced to a free field theory~\cite{martin,kapustin}. Later it was
shown~\cite{blasi} that in the abelian Chern-Simons theory with scalar fields
 $\be$-function also vanishes.

In our paper \cite{kaprecent} we proposed a proof of the nonrenormalization
 theorem for the gauge coupling in the theory of spinors interacting with the
 Chern-Simons field. The proof was based on the study of the conformal anomaly
 and resembled that in~\cite{blasi}.
 Here we want to extend the arguments of~\cite{kaprecent} to the general
 renormalizable $2+1D$ theory of scalars and spinors interacting with each
other
 and with the gauge field. The gauge field action is assumed to be the
 Chern-Simons action or the topologically massive action. The proof is valid in
 both abelian and nonabelian cases. The paper is organised as follows. In
 Section 2 we derive certain useful expressions for the trace of the
 renormalized energy-momentum tensor and use them to calculate the conformal
 anomaly in our theory. In Section 3 vanishing of $\be$-function is proved for
 the case when the gauge field action is the Chern-Simons action. In Section 4
 we prove that in the topologically massive gauge theory with matter fields
both
 $\be$-function of the gauge coupling and the anomalous dimension of the gauge
 field vanish.
\section{Conformal anomaly in Chern-Simons theory with matter fields.}
There is an intimate connection between the conformal anomaly and RG-functions.
Indeed, it is well known that the trace of the energy-momentum tensor can be
 expressed through the scale variation of the action:
\begin{equation}
\int T^\m_{\m} dx=\de S - \sum _i d_i \int \frac{\de S}{\de \phi _i}
dx, \label{eq:clastrace}
\end{equation}
where $\de S=\frac{d}{dt} \left| _{t=1} S[\phi ^t]\right.$, $\phi _i^t
(x)=t^{d_i}\phi _i
(tx)$, $d_i$ is the canonical dimension of~$\phi _i$.
The quantum analogue of this formula is
\begin{equation}
\left\langle \int T_{ren  \m}^\m dx\right\rangle _j=\de \Ga _{ren}
+\sum _i d_i\int j_i \phi _i dx, \label{eq:quantrace}
\end{equation}
where $j_i =-\de \Ga _{ren}/\de \phi_i$. Renormalization alters
 the scaling properties of $\Ga $ which results in the appearance of the
anomalous part in the renormalized trace of the energy-momentum tensor. So one
can hope to express the conformal anomaly through RG-functions. A well known
example is a formula for conformal anomaly in $4D$ QCD derived in
{}~\cite{nielsen,duncan} :
\begin{equation}
T_{ren \m}^\m =\frac{2\be (g)}{g} \frac{1}{4} \left( F_{\m\n} F^{\m\n}
 \right) _{ren}+(1+\de (g))m\!\left( \bar{\psi}\psi \right) _{ren}.
\label{eq:4Dtrace}
\end{equation}
In general the behaviour of $\Ga_{ren}$ under scale transformation is
 determined by RG equations. As is known their explicit form depends on the
 subtraction scheme. Suppose we consider a theory parametrized with $N$
 dimensionless couplings $g_1,\ldots,g_N$ and a mass $m$. We will use a mass
independent subtraction scheme, i.e., we will make all subtractions at
 $p^2=\m^2$. Then standard RG arguments~\cite{zinn-justin} lead to the
 following homogeneous RG equation (first derived by Weinberg~\cite{weinberg}):
\begin{equation}
\de\Ga_{ren} =\sum_{a=1}^N \be_a\frac{\partial\Ga_{ren}}{\partial
 g_a} -(\de +1)m\frac{\partial\Ga_{ren}}{\partial m} +\sum_i\ga_i\int\phi_i j_i
\,dx \label{eq:RGequation}
\end{equation}
Here $\be_a,\ga_i,\de$ depend not only on $g_1,\ldots,g_N$, but also on
 $m/\m$~\cite{weinberg} and are defined as
\begin{equation}
\be_a=g_a\m\frac{\partial}{\partial\m}\ln Z_a,\quad
 \ga_i=\frac{1}{2}\m\frac{\partial}{\partial\m}\ln Z_i,\quad
\de=\m\frac{\partial}{\partial\m}\ln Z_m \label{eq:RGfunctions}
\end{equation}
Substituting ~(\ref{eq:RGequation}) into ~(\ref{eq:quantrace}) we get
\begin{equation}
\left\langle \int T_{ren  \m}^\m dx\right\rangle _j=
\sum_{a=1}^N \be_a\frac{\partial\Ga_{ren}}{\partial
 g_a} -(\de +1)m\frac{\partial\Ga_{ren}}{\partial m}
+\sum_i\left(\ga_i+d_i\right)\int\phi_i j_i \,dx \label{eq:basic}
\end{equation}

Let us use~(\ref{eq:basic}) to compute conformal anomaly in the most general
 renormalizable theory of scalars and spinors interacting with the Chern-Simons
 field:
\begin{eqnarray}
{\Lc}=&\frac{1}{2} \ep^{\m\n\r} \left( A^a_\m\partial_\n A^a_\r +\frac{1}{3}
gf^{abc}A^a_\m A^b_\n A^c_\r\right) +b\partial_\m
 D^\m c+B\partial^\m\! A_\m\nonumber\\
 &\mbox{}+\left( D_\m \f \right) ^\da \!D^\m \f
 -m_\f^2\f^\da\f +\bar\psi\left( i\!\not\!\! D-m\right)\psi +
 {\Vc}(\psi,\f).
\label{eq:lagrcs}
\end{eqnarray}
We chose Landau gauge for definiteness. The most general potential $\Vc$
allowed by renormalizability is
\begin{equation}
{\Vc}=a_1 \bar\psi\psi\f +a_2 \bar\psi\psi\f\f +a_3\f^3+
a_4\f^4+a_5\f^5+a_6\f^6.\label{eq:V}
\end{equation}
Here $\bar\psi\psi\f$ is a trilinear invariant linear in $\bar\psi,\psi$
 and$\f$, $\f^3$ is a trilinear invariant built of $\f$ and
 $\f^\da$ and so on. The existence of such invariants depends on the
gauge group and the matter field representations.

The theory described by the Lagrangian~(\ref{eq:lagrcs}) is renormalizable by
 power counting provided  gauge anomaly is absent. To show the absence of the
anomaly it suffices to exhibit a BRST-invariant regularization scheme.
 Evidently higher covariant derivatives supplemented by the Pauli-Villars
 regularization of the one-loop diagrams serve our aim. To apply
 formula~(\ref{eq:basic}) to our theory let us rescale the parameters of the
 Lagrangian~(\ref{eq:lagrcs}) in the following way:
\begin{equation}
m_\f^2=\alpha_0m^2,\qquad a_b=\alpha_b m^{dim\,a_b}\quad\hbox{for}\quad
 b=1,\ldots ,6.\label{eq:rescale}
\end{equation}
Now the Lagrangian is parametrized with eight dimensionless couplings
 $g$, $\alpha_0,\ldots\alpha_6$ and a mass $m$, just as required for
 formula~(\ref{eq:basic}) to be applicable. The renormalized Lagrangian is
\[{\Lc}_{ren}=\frac{Z_3}{2} \ep^{\m\n\r} \left( A^a_\m\partial_\n A^a_\r
+\frac{1}{3}g\frac{\tilde Z_1}{\tilde Z_3}f^{abc}A^a_\m A^b_\n
 A^c_\r\right) +\tilde Z_3b\,\partial_\m D_{ren}^\m c+B\partial^\m\ A_\m\]
\[\mbox{}+Z_{2\f}\left(\left( D_{ren\,\m} \f \right) ^\da
 \!D_{ren}^\m \f -\bar Z_0Z_m^2\alpha_0m^2\f^\da\f\right)\]
\begin{equation}
 \mbox{}+Z_2\bar \psi \left( i\!\not\!\!D-Z_mm\right) \psi + {\cal
 V}_{ren}(\psi,\f).\label{eq:lagrren}
\end{equation}
Here $D_{ren}^\m$ is obtained from $D^\m$ by a change
 $g\rightarrow\frac{\tilde Z_1}{\tilde Z_3}g$ and ${\Vc}_{ren}$ is
\begin{eqnarray}
{\Vc}_{ren} & = &
\bar Z_1Z_mZ_2Z_{2\f}^{1/2}\alpha_1m^{1/2}\bar\psi\psi\f+
\bar Z_2Z_{2\f}\alpha_2\bar\psi\psi\f\f+
\bar Z_3Z_m^{3/2}Z_{2\f}^{3/2}\alpha_3m^{3/2}\f^3\nonumber\\
 & &\mbox{}+\bar Z_4Z_mZ_{2\f}^2\alpha_4m\f^4+
\bar Z_5Z_m^{1/2}Z_{2\f}^{5/2}\alpha_5m^{1/2}\f^5
+\bar Z_6Z_{2\f}^3\alpha_6\f^6.
\label{eq:Vren}
\end{eqnarray}
RG functions are defined according to
\[
\be =g\m\frac{\partial}{\partial\m}\ln\frac{Z_3^{1/2}\tilde Z_3}{\tilde Z_1}
,\qquad\be_a=\alpha_a\m\frac{\partial}{\partial\m}\ln\bar
 Z_a\quad\hbox{for}\quad a=0,\ldots ,6,\]
\[\de=\m\frac{\partial}{\partial\m}\ln Z_m,\qquad
\ga_A=\m\frac{\partial}{\partial\m}\ln Z_3^{1/2},\]
\begin{equation}
\ga_c=\ga_b=\m\frac{\partial}{\partial\m}\ln\tilde Z_3^{1/2},\quad
\ga_\psi=\m\frac{\partial}{\partial\m}\ln Z_2^{1/2},\quad
\ga_\f=\m\frac{\partial}{\partial\m}\ln Z_{2\f}^{1/2}.
  \label{eq:ourRGfunc}
\end{equation}
The expectation value of the zero momentum-transfer part of $T_{ren\,\m}^\m$ is
given by the general formula~(\ref{eq:basic}) where the first sum extends  over
all eight couplings $g,\alpha_0,\ldots,\alpha_6$. We can represent the
derivatives of $\Ga_{ren}$ entering~(\ref{eq:basic}) as expectation values of
renormalized operator insertions using rules of~\cite{lowenstein}. After this
we may remove the angular brackets since the sources $j$ are arbitrary and get
the following identity for renormalized operator insertions:
\begin{eqnarray}
\int T_{ren\,\m}^\m\,dx&=&\quad -\int\frac{\be}{g}\ep^{\m\n\r}
 \left( A^a_\m\partial_\n A^a_\r +\frac{1}{3} gf^{abc}A^a_\m A^b_\n
 A^c_\r\right)_{ren}dx\nonumber\\
& &\mbox{}+\int\left[\sum_{a=1}^6\be_a\left(\frac{\partial{\cal
 V}}{\partial\alpha_a}\right)_{ren}
-\be_0m^2\left(\f^\da\f\right)_{ren}\right] dx\nonumber\\
& &\mbox{}+(1+\de)\int\left(m\bar\psi\psi
+2m^2\alpha_0\f^\da\f-m\frac{\partial{\Vc}}{\partial
 m}\right)_{ren} dx\nonumber\\
&&\mbox{}+\sum_iA_i\int\left(\phi_i\frac{\de S}{\de\phi_i}\right)_{ren}dx,
\label{eq:3Dtrace}
\end{eqnarray}
where $A_i$ are finite coefficients which can be easily computed (for our aims
their concrete values are not important). This formula is a $2+1D$ analogue
 of~(\ref{eq:4Dtrace}).
\section{Nonrenormalization theorem for the Chern-Simons coupling.}
In this Section we are going to show that one of the consequences
 of~(\ref{eq:3Dtrace}) is $\be=0$. First let us find the most general form of
$T_{ren\m}^\m$ following the approach of~\cite{duncan}. To this end we
 consider the regularized trace $T_{reg\m}^\m$. Since the regularized action
is a sum of the classical action and the regulator terms, the regularized trace
contains a classical piece and a piece coming from regulator terms. The
 classical part of $T_{reg\m}^\m$ is obviously an operator of dimension 3 and
ghost number 0. Moreover, it is a gauge invariant operator up to terms of the
form $\phi_i\frac{\de S}{\de\phi_i}$. A well-known result of the theory of
operator mixing is that after regularization such an operator can mix only with
local operators of the same ghost number and the same or lower dimension. It is
 shown in~\cite{lee} that BRST-invariance restricts further the set of
operators
 which can mix, namely a gauge invariant operator of ghost number 0 can mix
only
 with gauge invariant operators and operators of the form $\phi_i\frac{\de
 S}{\de\phi_i}$. (The proof uses only BRST-invariance and antighost equation
of motion, therefore it is valid in any gauge field theory provided the gauge
anomaly is absent. It was shown above that this is the case for $2+1D$ gauge
theories.) One can easily see that the same is true for the ``regulator'' part
of $T_\m^\m$. Indeed the regulator term consists of the Pauli-Villars-ghost
term and the higher-covariant-derivative term. They are both gauge invariant by
definition and have ghost number 0. The Pauli-Villars term has dimension 3. The
higher-covariant-derivative term has dimension $d$ greater than three. However,
it is multiplied by $\Lambda^{3-d}$ ($\Lambda$ is the cutoff), hence it is an
irrelevant operator in RG-terminology~\cite{zinn-justin}, that is in the limit
$\Lambda\rightarrow\infty$ it is reduced to a linear combination of operators
of dimension 3 and lower. Hence the ``regulator'' term in $T_{reg\m}^{\m}$
 also mixes only with gauge invariant operators of dimension 3 and lower and
 with  operators of the form  $\phi_i\frac{\de S}{\de\phi_i}$.

One can see that there are only eight linearly independent gauge invariant
 operators with ghost number 0 and dimension 3 or lower which are not reduced
to  $\phi_i\frac{\de S}{\de\phi_i}$ with some $i$. These are
\[O_0=\f^\da\f,\quad O_1=\bar\psi\psi\f,\quad
 O_2=\bar\psi\psi\f\f,\quad O_3=\f^3,\]
\begin{equation}
 O_4=\f^4,\quad O_5=\f^5,\quad O_6=\f^6,\quad O_7=\bar\psi\psi.
\label{eq:Loperators}
\end{equation}
These are just the operators out of which the Lagrangian~(\ref{eq:lagrcs}) is
 built. However, the Chern-Simons density does not appear here (though it
 appears in~(\ref{eq:lagrcs})) because it is not gauge invariant. So
 $T_{ren\m}^\m$
 is a linear combination of $(O_i)_{ren}\,,i=0,\ldots,7$. Using this
information
we can reconstruct $T_{ren\m}^\m$ from its zero momentum-transfer part found
in Section 2 just as it was done in~\cite{duncan}. However here an important
difference arises as compared to the $3+1D$ case. Namely~(\ref{eq:3Dtrace})
contains a term (the Chern-Simons term) which cannot be present in
 $T_{ren\m}^\m$ because it is not one of $O_i$\,. Hence its coefficient
 $\be$ must be zero. This is the desired result. From the
 definition~(\ref{eq:ourRGfunc}) it is clear that this means that $Z_g$ is
finite. Hence the corresponding Callan-Simanzik function also vanishes. (Our
$\be$ is {\it not} the Callan-Simanzik function because we chose a
 mass-independent subtraction scheme. Neither is it the Gell-Mann--Low function
because our $m$ does not determine the physical pole of the propagator.)
\section{Topologically massive gauge theory.}
The case of the topologically massive gauge theory is much simpler. We add to
the Lagrangian~(\ref{eq:lagrcs}) the Maxwell term $1/M Tr F_{\m\n}^2$ where
 $M$ is a new mass parameter. Power counting shows that the gauge field
 self-energy diverges at most linearly and the 3-gluon vertex diverges at most
logarithmically. Hence no counterterm of the form $Tr F_{\m\n}^2$ is
 needed. Then $Z_1/Z_3$ must be finite. Indeed, let $Z_{M3}$ and $Z_{M1}$ be
 the renormalization constants for the 2- and 3-gluon Maxwell vertices
 correspondingly. Ward identities require that $Z_1/Z_3=Z_{M1}/Z_{M3}$. But we
have just seen that $Z_{M1}$ and $Z_{M3}$ are finite, therefore $Z_1/Z_3$ is
 finite and $\be =2\ga_A$.

Now let us show that $Z_3$ is finite. The Schwinger-Dyson equation for the
gluon
self-energy is represented graphically on Fig.~1. We are interested only in the
 part of the gluon self-energy which has superficial degree of divergency $
\omega=1$. Gauge invariance tells us that the divergent piece is proportional
to $\ep^{\m\n\r}p_\r$. But in the ultraviolet the irreducible vertices
 in these diagrams have the same form as the tree vertices apart from
 logarithmic factors (Weinberg's theorem and Lorenz-invariance), therefore they
 are P-even (proportional to $\ga_\m$ in the case of the $\bar\psi\psi A$
 vertex). The same is true about exact propagators; in particular the gluon
 propagator has Maxwell asymptotics. Hence the parts of the diagrams which
have $\omega=1$ are P-even (contain even number of $\ga$-matrices in the case
of diagrams (b) and (d)) and cannot give the divergency proportional to
 $\ep^{\m\n\r}p_\r$. We conclude that all these diagrams are in fact
finite and hence $Z_3$ is finite and $\be=\ga_A=0$ (essentially the same
argument was used in~\cite{rao} to prove the ultraviolet fineteness of
 the topologically massive gauge theory without matter fields).

We showed that the gauge coupling $g$ is not renormalized both in the
 Chern-Simons gauge theory and in the topologically massive gauge theory. In
the topologically massive case the reason for this is just the presence of the
Maxwell term which serves as an ultraviolet regulator, so the gauge field part
of the action is not renormalized. In the Chern-Simons theory with matter
fields
$\be$-function vanishes because it turns out to be a coefficient which
determines the mixing between the trace of the energy-momentum tensor and the
Chern-Simons density. This mixing is in fact forbidden because the Chern-Simons
density is not a gauge invariant operator, therefore $\be$ must be zero. In
particular vanishing of $\be$ means that the global anomaly is absent.

\newpage
\begin{figure}

\setlength{\unitlength}{0.6pt}
\input FEYNMAN
\textheight 800pt \textwidth 450pt
\begin{picture}(18000,18000)(0,-6000)
\put(-2000,20000){$\Large\Pi$} \put(0,20000){$\Large =$}
\multiput(3000,6000)(0,-14000){2}{$\Large +$}
\drawloop\gluon[\W 5](10000,18000)                           
\global\Xone=\gluonbackx
\global\advance\Xone by -200
\global\Yone=\gluonbacky
\global\advance\Yone by \loopfronty
\global\divide\Yone by 2
\global\Ytwo=\gluonbacky
\global\Yfour=\Ytwo
\global\advance\Ytwo by -\Yone
\global\advance\Xone by -\Ytwo
\put(\Xone,\Yone){\circle*{500}}
\drawline\gluon[\W\FLIPPED](\Xone,\Yone)[2]
\global\Ythree=\loopfronty
\global\advance\Ythree by -1000
\global\multiply\Ytwo by 2
\multiput(\loopfrontx,\Ythree)(0,\Ytwo){2}{\framebox(2300,2000)}
\global\advance\Ythree by -3000
\put(\loopfrontx,\Ythree){\Large (a)}
\global\advance\loopfrontx by 2300
\drawline\gluon[\E\REG](\loopfrontx,\Yfour)[2]
\drawline\gluon[\E\FLIPPED](\loopfrontx,\loopfronty)[2]
\global\advance\gluonbacky by -1000
\global\advance\Yfour by -\loopfronty
\global\advance\Yfour by 2000
\put(\gluonbackx,\gluonbacky){\framebox(2000,\Yfour)}
\global\advance\gluonbackx by 2000
\drawline\gluon[\E\REG](\gluonbackx,\Yone)[2]            
\global\advance\gluonbackx by 3000
\multiput(\gluonbackx,20000)(0,-14000){3}{$\Large +$}
\global\advance\gluonbackx by 3000
\drawline\gluon[\E\REG](\gluonbackx,\gluonbacky)[2]      
\global\Xfive=\gluonbackx
\global\Yfive=\gluonbacky
\put(\Xfive,\Yfive){\circle*{500}}
\global\advance\gluonbackx by 2000
\put(\gluonbackx,\Yfive){\oval(4000,4000)[l]}
\global\Ythree=\Yfive
\global\advance\Ythree by -3000
\multiput(\gluonbackx,\Ythree)(0,4000){2}{\framebox(2300,2000)}
\global\advance\Ythree by -3000
\put(\gluonbackx,\Ythree){\Large (b)}
\global\advance\gluonbackx by 2300
\global\advance\Ythree by 4000
\drawline\fermion[\E\REG](\gluonbackx,\Ythree)[2000]
\global\advance\Ythree by 4000
\drawline\fermion[\E\REG](\gluonbackx,\Ythree)[2000]
\global\advance\fermionbacky by -5000
\put(\fermionbackx,\fermionbacky){\framebox(2000,6000)}
\global\advance\fermionbackx by 2000
\drawline\gluon[\E\REG](\fermionbackx,\Yfive)[2]           
\drawloop\gluon[\W 5](10000,4000)                          
\global\Xone=\gluonbackx
\global\advance\Xone by -200
\global\Yone=\gluonbacky
\global\advance\Yone by \loopfronty
\global\divide\Yone by 2
\global\Ytwo=\gluonbacky
\global\Yfour=\Ytwo
\global\advance\Ytwo by -\Yone
\global\advance\Xone by -\Ytwo
\put(\Xone,\Yone){\circle*{500}}
\drawline\gluon[\W\FLIPPED](\Xone,\Yone)[2]
\global\Ythree=\loopfronty
\global\advance\Ythree by -1000
\multiput(\loopfrontx,\Ythree)(0,\Ytwo){3}{\framebox(2300,2000)}
\drawline\gluon[\W\FLIPPED](\loopfrontx,\Yone)[2]
\global\advance\Ythree by -3000
\put(\loopfrontx,\Ythree){\Large (c)}
\global\advance\loopfrontx by 2300
\drawline\gluon[\E\REG](\loopfrontx,\Yfour)[2]
\drawline\gluon[\E\REG](\loopfrontx,\Yone)[2]
\drawline\gluon[\E\FLIPPED](\loopfrontx,\loopfronty)[2]
\global\advance\gluonbacky by -1000
\global\advance\Yfour by -\loopfronty
\global\advance\Yfour by 2000
\put(\gluonbackx,\gluonbacky){\framebox(2000,\Yfour)}
\global\advance\gluonbackx by 2000
\drawline\gluon[\E\REG](\gluonbackx,\Yone)[2]              
\global\advance\gluonbackx by 6000
\drawline\gluon[\E\REG](\gluonbackx,\gluonbacky)[2]        
\global\Xfive=\gluonbackx
\global\Yfive=\gluonbacky
\put(\Xfive,\Yfive){\circle*{500}}
\global\advance\gluonbackx by 2000
\put(\gluonbackx,\Yfive){\oval(4000,5000)[l]}
\global\Ythree=\Yfive
\global\advance\Ythree by -3500
\multiput(\gluonbackx,\Ythree)(0,2500){3}{\framebox(2300,2000)}
\global\advance\Ythree by -3000
\put(\gluonbackx,\Ythree){\Large (d)}
\drawline\gluon[\W\FLIPPED](\gluonbackx,\Yfive)[2]
\global\advance\Xfive by 4300
\global\advance\Ythree by 4000
\drawline\gluon[\E\REG](\Xfive,\Yfive)[2]
\global\Xfour=\gluonbackx
\global\advance\Xfour by -\Xfive
\drawline\fermion[\E\REG](\Xfive,\Ythree)[\Xfour]
\global\advance\Ythree by 5000
\drawline\fermion[\E\REG](\Xfive,\Ythree)[\Xfour]
\global\advance\fermionbacky by -6000
\put(\fermionbackx,\fermionbacky){\framebox(2000,7000)}
\global\advance\fermionbackx by 2000
\drawline\gluon[\E\REG](\fermionbackx,\Yfive)[2]           
\global\advance\loopfrontx by -2300                        
\put(\loopfrontx,-5500){\framebox(2300,2000)}
\global\advance\loopfrontx by 2300
\drawloop\gluon[\SE 2](\loopfrontx,-4500)
\global\Xone=\gluonbackx
\global\advance\Xone by -\loopfrontx
\global\advance\loopfrontx by -2300
\global\advance\loopfrontx by -\Xone
\global\Yone=\gluonbacky
\drawline\gluon[\SW\REG](\gluonbackx,\gluonbacky)[2]
\drawloop\gluon[\N 2](\loopfrontx,\Yone)
\drawline\gluon[\SE\FLIPPED](\loopfrontx,\Yone)[2]
\global\Xtwo=\gluonbackx
\global\advance\Xtwo by 250
\global\Ytwo=\gluonbacky
\put(\Xtwo,\Ytwo){\circle*{500}}
\drawline\gluon[\W\FLIPPED](\Xtwo,\Ytwo)[3]
\drawline\gluon[\E\REG](\Xtwo,\Ytwo)[3]                    
\put(\Xfive,-6000){\oval(2000,2500)[tr]}                   
\global\advance\Xfive by -2300
\put(\Xfive,-6000){\oval(2000,2500)[tl]}
\put(\Xfive,-5500){\framebox(2300,2000)}
\global\Xsix=\Xfive
\global\advance\Xsix by 1150
\global\advance\Xfive by -1000
\put(\Xfive,-6000){\line(1,-2){2150}}
\global\advance\Xfive by 4300
\put(\Xfive,-6000){\line(-1,-2){2150}}
\put(\Xsix,\Ytwo){\circle*{500}}
\drawline\gluon[\W\FLIPPED](\Xsix,\Ytwo)[3]
\drawline\gluon[\E\REG](\Xsix,\Ytwo)[3]                    
\global\advance\Ytwo by -3000
\global\advance\Xtwo by -1100 \global\advance\Xsix by -1100
\put(\Xtwo,\Ytwo){\Large (e)}
\put(\Xsix,\Ytwo){\Large (f)}
\end{picture}
\vskip 150pt
\caption{Schwinger-Dyson equation for gluon self-energy.  Boxes denote exact
propagators and irreducible vertices. Curly lines correspond to gluons and
solid lines correspond to ghosts and matter fields.}
\end{figure}
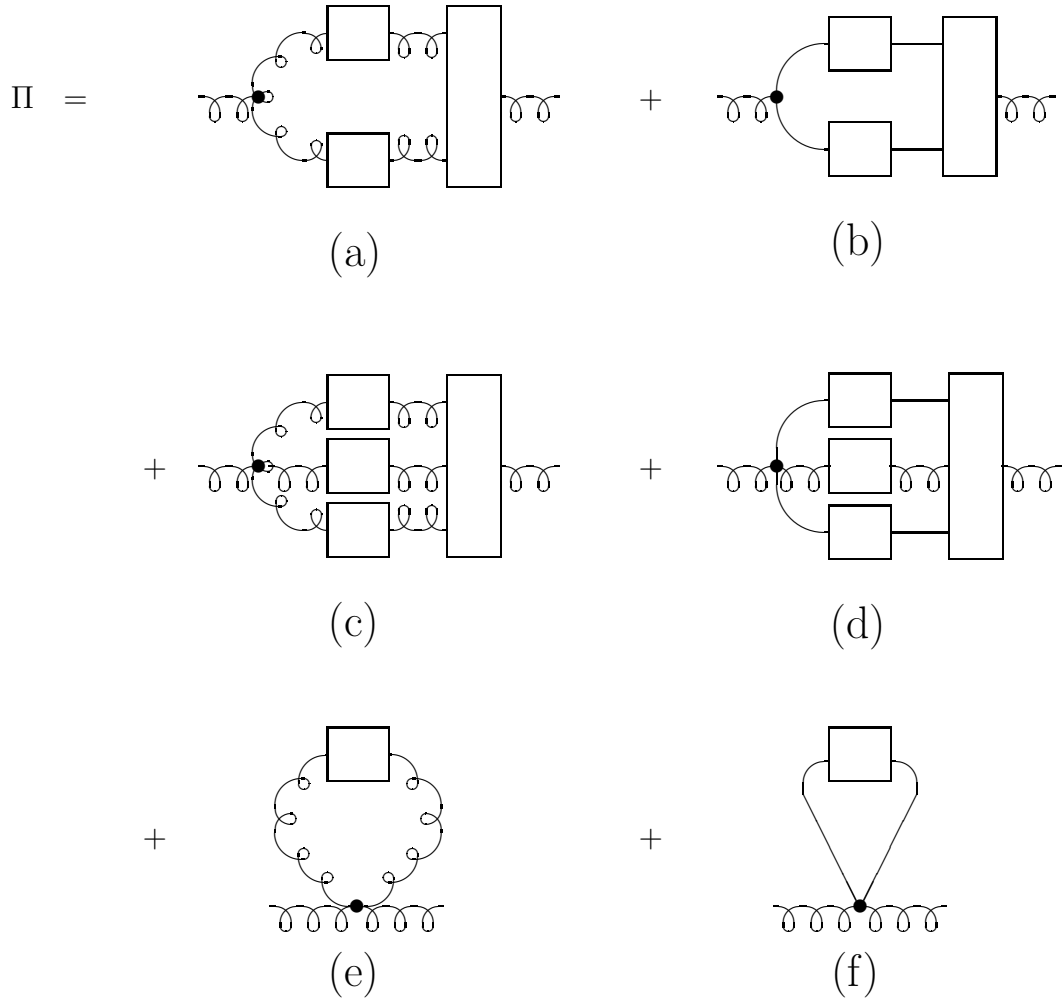

\end{document}